\documentclass[runningheads]{llncs}
\usepackage{graphicx}
\usepackage{color}
\usepackage{graphicx}
\usepackage{subfigure}
\usepackage{url}
\usepackage{array,multirow}
\usepackage[T1]{fontenc}
\usepackage{xcolor}
\usepackage{colortbl}
\usepackage{textcomp}
\usepackage{amsmath}
\newcommand{\fig}[1]{Fig.~\ref{#1}}

\begin{document}
\title{Authenticated Preambles for Denial of Service Mitigation in LPWANs}
%
%
\author{Ioana Suciu\inst{1}
\and
Jose Carlos Pacho\inst{1}
\and
Andrea Bartoli\inst{1}
\and
Xavier Vilajosana\inst{1,2}}

\authorrunning{I. Suciu et al.}
%
\institute{Worldsensing S.L, Viriat 47, Barcelona, Catalonia, Spain 
\email{\{isuciu,jcpacho,abartoli,xvilajosana@worldsensing.com\}}\\
\and
Universitat Oberta de Catalunya, Barcelona, Catalonia, Spain\\
\email{\{xvilajosana\}@uoc.edu}}
\maketitle              
\begin{abstract}
In this article we introduce authentication preambles as a mechanism to mitigate battery exhaustion attacks in LPWAN networks. We focus on the LoRaWAN technology as an exponent of industrial LPWANs. We analyze the impact of DoS attacks in Class B deployments and implement authentication preambles to limit attacker options when forcing nodes to overhear class B beacons. The article presents realistic results demonstrating significant energy savings (91\% energy saving when a network is attacked) versus a 4\% energy overhead of the mechanism in normally operating networks. 

\end{abstract}
\section{Introduction}

In the context of the new market demands, such as smart cities, e-health, intelligent transportation, infrastructure monitoring and many other industrial applications, wireless sensor networks allow users to remotely access data and take decisions based on it. For example, an infrastructure monitoring application could trigger alarms when a maintenance cycle is needed, allowing for taking timely and appropriate actions. 

Low Power Wide Area Networks (LPWANs), are wireless sensor networks communication technologies that provide low power operation and long range connectivity at the cost of reduced data rates and strict duty cycle regulations. Current technologies \cite{SemtechWhatIsLoRa,sigfox,weightless}, operate in the sub-GHz bands in order to cover a communication range in the order of kilometers, have a single hop network topologies and an Aloha or CSMA-based MAC access. 
Moreover, a gateway is able to accommodate thousand of sensor nodes, allowing for low cost deployments and customizable applications and services. 
The sensor nodes hardware and software is built to be simple and minimalistic with the goal to ensure years of battery lifetime. The low energy consumption and the possibility of powering the nodes using energy harvesting, reduces even more the costs of batteries and maintenance. All these factors make LPWAN the technology with the lowest energy consumption per provided service, but also makes them be vulnerable in front of attackers.

In \cite{cipsec_attacks}
the main cyberattacks faced by the critical infrastructure owners and operators are introduced. Amongst others, those wireless devices are exposed to phishing, unpatched vulnerabilities and Denial of Service (DoS) attacks. Phishing opens the door to a wide range of possibilities: traffic capture, network flood, controlling of network parameters and other further possible exploitations. In mesh networks, wormhole attacks are used to create false route information and routing loops that increase the energy consumption of these networks \cite{aras17selectivejamming}.

The eventual vulnerabilities existent in the application or operating system allow an attacker to perform actions for which it's not authorized and it mainly leads to collection of information. For example, in the case of an energy management system, an attacker could get information about when and where power is used, that could further lead to knowing if and when anyone is in that property~\cite{en_sabotage}.

Both phishing and the exploit of vulnerabilities can lead to DoS. The DoS prevents a system from carrying its designated tasks. Jamming can be a way towards service disrupt \cite{en_sabotage}. Moreover, in LPWANs, because of the low data rate that leads to a long time on air of the messages, jamming is possible and effective \cite{aras17selectivejamming}. 

In this paper, our attention is focused on the LoRa technology, one of the most used industrial LPWAN technology.
Security issues have been analyzed for LoRa networks~\cite{aras17selectivejamming}, \cite{tomasin17secAnalysisLora}, \cite{singh16iotsec};
Still, the analysis made in the literature explores how these issues impact the traffic performance and the actual data privacy, while in this work we focused on the impact on energy consumption and network lifetime. 
We analyze the impact of a DoS attack on the network lifetime through real data collection using the Loadsensing sensor nodes developed by Worldsensing \cite{worldsensing}. The novelty of this work consists in considering LoRaWAN class B in a scenario in which an attacker aims at draining the batteries of the sensor nodes in order to kill the network. Then we evaluate the efficiency of a possible solution based on authenticated preambles against this type of attack.

\section{Security mechanisms in LoRaWAN}

LoRaWAN \cite{SemtechWhatIsLoRa} is a promising technology for IoT. It's proprietary physical layer uses CSS modulation\cite{Semtech13LoRa}. Orthogonal spreading factors (SF) allow for variable data rates ranging from 0.3 kbps to 27 kbps. SF can vary from 7 to 12, the least corresponding to the smallest datarate and highest communication range. 
LoRa enables the trade-off between throughput for coverage range and robustness while keeping a constant bandwidth.
In Europe, the sensor nodes can send data on randomly chosen channels in the 868MHz ISM band, subject to the allowed duty cycle \cite{etsi-rule,limitslora}. A typical gateway can listen on 8 channels at once. 

LoRaWAN follows a star topology: the end-devices or sensor nodes communicate directly with a LoRa gateway. There are three categories of end-devices \cite{lorawan_spec}: Class A, Class B and Class C, but for all devices is mandatory to be able to support class A by default. A class A end-device supports bi-directional communication, in the sense that a DL transmission can be received only in the pre-defined reception windows that the device opens following it's UL data. Class A devices provide the lowest possible energy consumption: the end-device transmits messages using Aloha protocol restricted by a mandatory 1\% duty cycle \cite{etsi-rule}. Normally, no acknowledgements are provided by the gateway, as these are
expensive in terms of energy. The DL traffic is the mainly dedicated for MAC commands for making an end-device use a different datarate, channel or transmission power \cite{lorawan_spec}. 

Class B end-devices have additional receive windows determined by the gateway's beaconing interval (1s to 128s). Class C devices allow continuous
reception of data. Industrial solutions based on LoRaWAN use class A devices. Class B devices would allow for more feedback from the gateway.

Because of the Aloha-based protocol, collisions of signals can happen at the gateway. Collisions happen if two or more packets are sent on the same channel, with the same SF and they overlap in time. In case of collision, all of the collided packets are dropped. For two packets arriving at the gateway with the same SF at the same time, the gateway could decode one if it has a power greater than 6dB above the other peak \cite{georgiou17lorascale}.
As for different SFs the rejection gain ranges from 16 to 36 dB, we can consider there is no inter-spreading factor interference.

In what concerns non-LoRa interferers, due to the redundancy associated with wideband spread-spectrum modulation, LoRa is  resilient to the interference mechanism that
appears as bursty short duration pulses \cite{SemtechModBasics}.
According to \cite{kamil18loraperformance}
, LoRa can tolerate a non LoRa interferer if this is less than 5dB (19.5dB) above desired signal for SF=7 (12) for the case of an error coding scheme of 4/6. Being wide-band, a narrow band jamming signal would only add noise on a very small portion of this band and the LoRa signal would still be recoverable. Also, a jammer that floods the channel can easily be detected and dealt with. Moreover, it would need to transmit with high energy on a very wide band of the radio spectrum, which poses an important problem for a potential attacker.

LoRaWAN offers a good degree of protection against impersonation, as the end-device needs to be authenticated: a message authentication code (MAC) confirms that the message comes from an authorized sender. The LoRaWAN network and application layer use EUI64, while the device specific key, EUI128. AES CCM (128-bit) is used for encryption and
authentication \cite{lorawan_spec}.
The network session key (NwkSKey) is used for checking the validity of messages (MIC- message integrity check). 
The application session key (AppSKey) is used for encryption and decryption of the payload.

Regarding replay attacks, LoRaWAN offers a mechanism
to prevent them \cite{aras17selectivejamming}: the MIC of a message, once validated by
a gateway, prevents any further occurrences of the same sequence number. The lack of timestamp in LoRaWAN headers, makes it possible for a packet to be replayed at a later time as legitimate, only if the original message was jammed and no message with a higher sequence number has been received by the gateway. This attack could be used to hide the changes detected by the sensor nodes.

A more subtle type of attack that can be classified as a denial of service is represented by the exhaustion attacks: it exploits the communication protocol in order to drain the battery of the  device. 
The lack of authentication at link or network layer can be exploited by injecting forged packets in the network. Also, another way would be by making a given sensor node continuously transmit and receive messages. In the former case, the malicious packet is often detected at the application layer and thus precious network resources are wasted processing the packet. In the latter case, an attacker sending useless messages can exhaust the node's resources, as these messages are completely received before being discarded. This type of attack is unpractical for the case of class A devices, as the only opportunity it has is after the uplink transmission, which happens normally every few hours. Class B devices have a higher listening rate and so are much more exposed to this type of attack, especially on devices that require high reactivity.

\section{Preamble Authentication in LoRaWAN}
\subsection{Exhaustion attacks in LoRaWAN}
In the industrial context, sensor networks must be resilient and robust against any external disruption. We address the problem of exhaustion attacks, which is a type of Denial of Service (DoS) attack for battery powered devices. This type of attack exploits the communication protocol in order to drain the device battery, rendering it inoperative.
We consider LoRaWAN class B end-devices, as class A devices have a reactivity limited to the transmission rate of the uplink messages, although the attack still applies.  Class B allows for an efficient downlink communication at the expense of increased power draw due to periodic listening for beacons.

To carry out this attack, an attacker sniffs the medium for any downlink message addressed to the target end-device (for class A a node may listen for the uplink message to be able to attack the downlink windows afterwards). Then the attacker would synchronize with the listening window of the LoRaWAN device and send a single but very long packet on each listening window. Since the device needs to receive the whole packet in order to calculate the network level message authentication code (MAC) before discarding it, it would be forced to receive up to 255 bytes of payload message which could take up to 14 seconds depending on the SF in use. 

\subsection{Early message authentication}
\begin{figure}[h]
  \subfigure[]{\includegraphics[width=\textwidth]{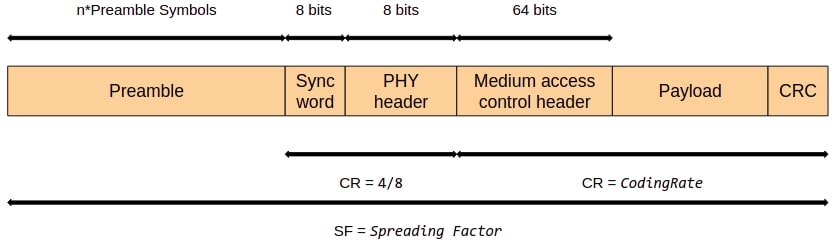}\label{before}}
  
  \subfigure[]{\includegraphics[width=\textwidth]{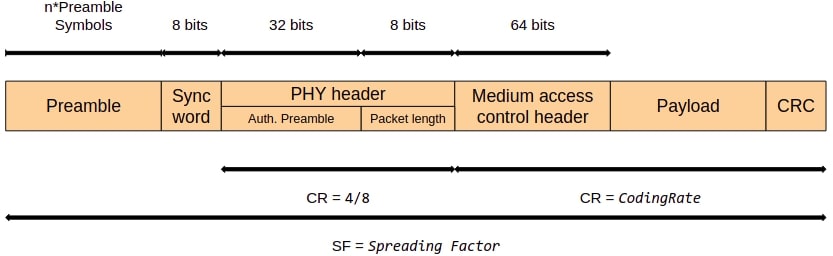}\label{after}}
  \caption{Packet structure: a) as defined by LoRaWAN; b)new structure allowing early authentication of the packets received by an end-device.}
  \label{fig:exph,exgr}
\end{figure}

We propose a verification method at the PHY-layer that is extensible to any wireless protocol. The use of an authentication preamble (AP) is able to reject malicious packets sooner, saving energy and so guaranteeing network's long-term availability.  \fig{before} shows the packet structure as used in LoRaWAN, while \fig{after} shows the proposed packet structure that would lead to a sooner verification of a message authenticity. This allows discarding the malicious packet after receiving the first 4 bytes after the synchronization word.

Normally, the message authentication code (MAC) is generated using the payload of the message it accompanies. As we want to be able to reject the packet sooner, we would not have the payload in order to compute the MAC. Therefore we must generate a MAC that is known at the start of the reception frame and that is different at each frame. We propose a token exchange scheme: the end-device uses a token that the gateway will use to authenticate at the physical layer all subsequent communication. 

\begin{figure}[h]
\includegraphics[width=\textwidth]{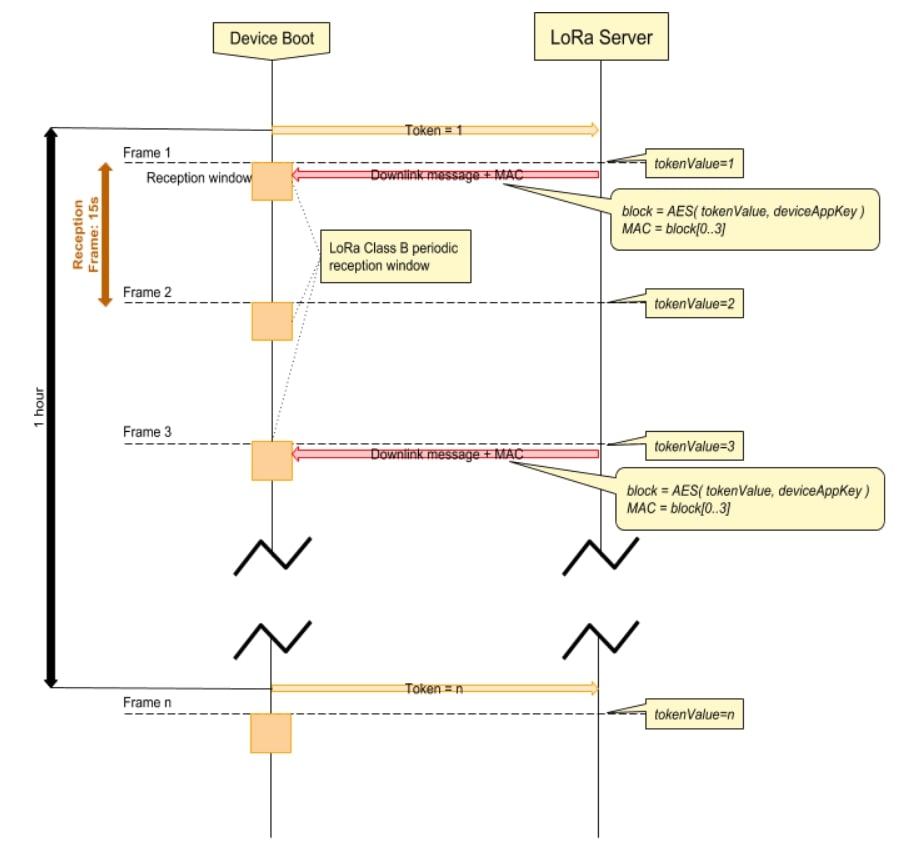}
\caption{Token exchange as a form of message authentication: the first token value is generated when the end-device is booted. The token value is then used to authenticate any packet coming from the LoRa server and through the gateway.} \label{f1}
\end{figure}

The end-device does not have to authenticate its messages to the gateway at the physical layer. This is because the gateway does not have energy constraints. In \fig{f1} we propose to use a frame counter as the token upon which the MAC will be generated.
Given that each reception frame has a fixed duration, the gateway, once it has obtained the frame counter from the node, can easily predict the frame counter no matter how much time has passed. If the gateway knows the counter value $\phi$ for the frame $f_i$, to know the counter for a frame in the future $f_m$ it simply needs to divide the elapsed time between $f_i$ and $f_m$ by the frame duration. The resulting value is then added to $\phi$ to obtain the frame counter for $f_m$.

\subsection{Securing the token exchange} \label{sec:MAC}
The token is sent to the gateway once the end-device boots. This ensures that reception is available as soon as possible. To deal with possible de-synchronization, the token is retransmitted periodically at a predefined interval. The token exchange must be coupled with a strong cipher algorithm such as AES. The application key is distributed during the device manufacturing and it is only known by the device itself and by the network server. A possible vulnerability arises when an adversary can capture the moment a device reboots because it will then know the exact token value. 
The token value can be initialized during manufacturing to a random value for each device, which would remove this vulnerability.

The message authentication code (MAC) is then calculated as follows:
\begin{enumerate}
\item The frame counter value $k$ is computed
\item The value $k$ is padded to up to 16 bytes. The 16 byte block is then encrypted using an AES algorithm with a shared key 
\item The last 4 bytes of the resulting cipher text represents the MAC value. 
\end{enumerate}

The end-device follows the same procedure in order to determine the accepted MAC value for the current reception frame. 

Regarding the suitability of our proposed approach in face of a brute force attack,  which is an attacker trying to sniff and guess the next MAC value, the 64-bit token makes the search space close to intractable, even if using rainbow tables \cite{rainbowtables}. Also, a cipher-text attack would result difficult because when using either a cipher or a hash on the token, the entire result of the computation, usually a 16 bytes block, is not sent over the air, only 4 bytes of it, in the MAC field. The adversary does not then have a complete set of cipher-text to attack the cipher algorithm.

\section{Evaluation}
\begin{figure}

\includegraphics[width=\textwidth]{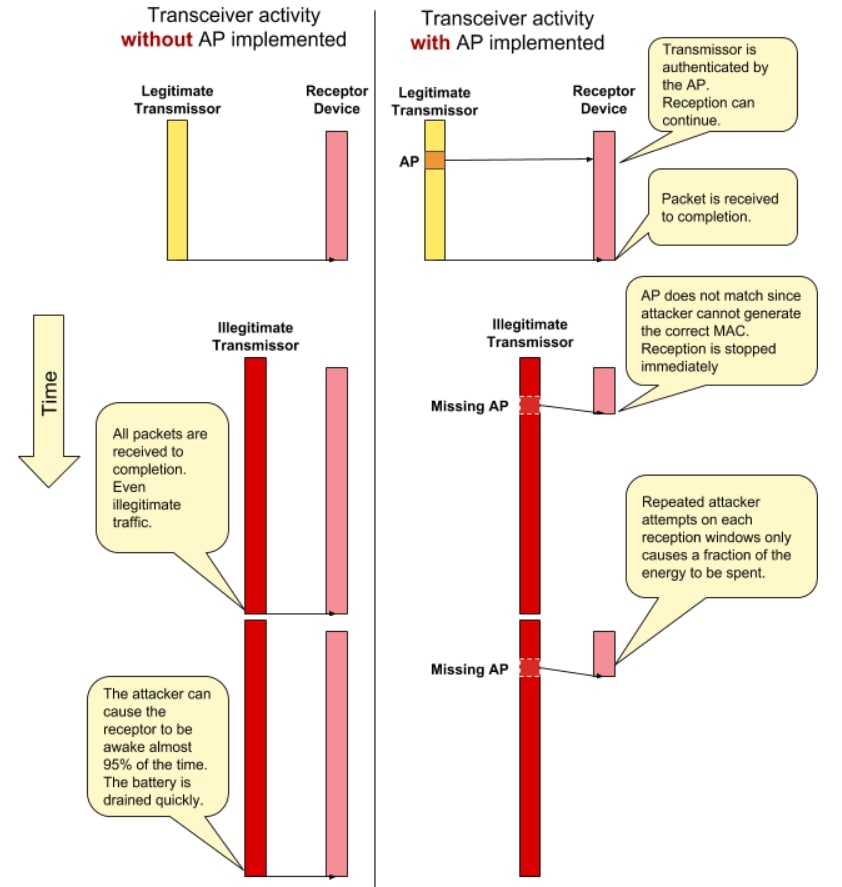}
\caption{The behavior of an end device subject to an energy exhaustion attack: (left) the end device does not implement AP; (right) end device implements AP. } \label{f2}
\end{figure}

In this section we evaluate the efficiency of using authentication preambles in a realistic LoRaWAN setting. \fig{f2} shows the behavior of an end device in both cases of using and not using authentication preamble. On the left side of the figure we can see that if the end device does not use AP it will stay in reception mode for all the packets sent by an attacker, without being able to discard them before their transmission is completed. On the right side of \fig{f2}, an end device implementing AP is able to discard the attacker's packet, 4B after the synchronization word composing this message. 

The experimental setup we used was composed of a Loadsensing sensor node \cite{loadsensing}, a LoRaWAN gateway and a sniffer that would synchronize with the node and it would send it packets, acting as if it was a legitimate gateway. The end device is implementing LoRaWAN class B and is set to wake up every 15s to listen for DL messages coming from the LoRaWAN gateway. The predefined listening period for downlink packets from the gateway is 300ms. The sensor node is configured to take samples at a very low rate (12h - 24h), and thus most of the time is sleeping.

In a typical scenario, the DL messages from the gateway are very limited, so the end device would normally wake up, listen for 300ms and go back to sleep, as there would be no packets to hear. This is why, in our setting, the attacker is the only one sending DL messages to the end device. The attacker would send every 15s (the wake-up period of the end device) a packet with the payload set to the maximum allowed value (242B)\cite{lorawan_spec}. Using SF12, this transmission takes 14s. The goal of the attacker is to keep the node awake as much time as possible. 

\subsection{Energy exhaustion attack: End device does not implement AP }
\begin{figure}
\includegraphics[width=\textwidth]{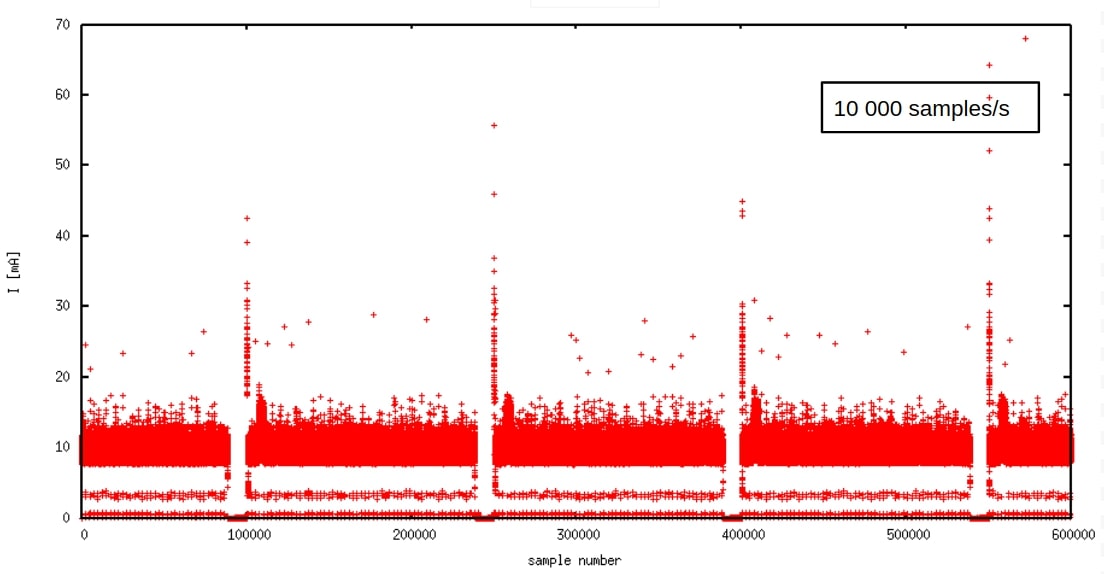}
\caption{{End device current consumption versus time (sample number), for the case when AP is not implemented and the attacker sends packets with the maximum allowed payload size. Listening periodicity: 15s; Packet duration: 14s; Awake time: 14s; Test duration: 1min.}} \label{test2}
\end{figure}

\fig{test2} shows the end device current consumption versus time, for the case when AP is not implemented and the attacker sends packets with the maximum payload size. 

We can observe the periodicity in node's activity: the node wakes up, 
starts receiving the packet and cannot discard it, it has to receive it completely before being able to see that it is not a legitimate packet. For obtaining this plot, we used the PowerScale tool \cite{powerscale}, taking measurements for 1 minute and repeating the tests 1000 times. There are 10 000 samples/s, so the whole test shows 600 000 samples. We can see in \fig{test2} that the packet reception lasts for approximately 14s out of the 15s configured as beaconing interval.

\subsection{Energy exhaustion attack: End device implements AP }

This setting is similar to the previous one, with the only difference that the end device implements the AP and it is expecting the gateway to send the correct message authentication code described in Section \ref{sec:MAC}.

\begin{figure}
\includegraphics[width=\textwidth]{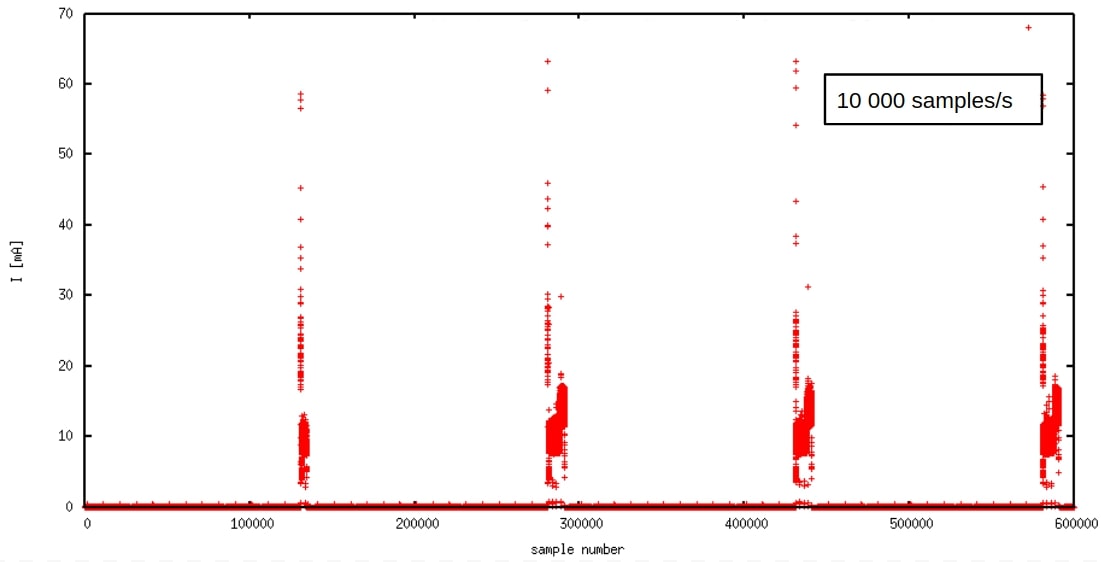}
\caption{End device current consumption versus time (sample number), for the case when AP is implemented and the attacker sends packets with the maximum allowed payload size. Listening periodicity: 15s; Packet duration: 14s; Awake time: 1s; Test duration: 1min.} \label{test3}
\end{figure}

In \fig{test3} we can see that when using AP, the node wakes up every 15s, but it is able to discard attacker's packet after checking for the existence of AP. As the attacker does not have any AP, or it is not able to generate the correct MAC, the end device is able to go back to sleep state.
Every 15s the node will wake up for approximately 1s.

\subsection{Analysis of Energy Consumption}
In this section we analyze the energy consumption of a node running in a typical Class B LoRaWAN network. The results have been extrapolated from real measurements conducted on a LoadSensing device\cite{loadsensing}. 
The performed tests aimed at understanding what is the energy drained from the battery of a node in normal operation conditions and under a DoS attack, when implementing the preamble authentication. The parameters we used for the analysis are presented in Table \ref{tab0}.

\begin{table}
\caption{Technical parameters }\label{tab0}
\begin{tabular}{p{5cm}p{3cm}p{4cm}}
\hline
Parameter &  Value & Unit\\
\hline
Battery Capacity & 23.2  & Ah\\
Sensor power draw & 2.2  & Ah/year\\ 
Sensors UL data  & 0.025 & Ah/year\\
DL listening & 1.983 & Ah/year\\
AP Beacon & 0.17 & Ah/year\\
Attack power drain (no AP) & 94.024 & Ah/year\\
Attack power drain (with AP) &  6.354 & Ah/year\\
\hline
\end{tabular}
\end{table}

We considered the real case of a Loadsensing end device using
4x Li-SOCl2 primary batteries of 5.8 Ah each. It sends a 20 B data message per day, transmission which takes about 5s in the air. The used radio has a transmission energy consumption (at 7dB) of 18mA. Reception energy consumption is 11.5mA. 

The sensor node wakes up every 15s to listen for DL data. In normal operation mode the listening duration is 300ms, while in the case of an DoS attack, if the node implements the AP, it can discard the packet after 0.9s. If no AP is implemented, the node stays in reception mode for 14s.

In Table \ref{tab1} the expected battery duration of the device is presented, considering a normal network operation and when a DoS attack is performed. Preamble authentication is considered incurring an extra overhead of 4B which causes extra energy consumption.


\begin{table}
\caption{Expected battery lifetime summary considering normal operation and exhaustion attacks for nodes using an authentication preamble. }\label{tab1}
\begin{tabular}{p{4cm}p{4cm}p{4cm}}
\hline
Operation Mode &  Authentication Preamble & Battery Duration\\
\hline
Normal Operation &  No & 5.51 years\\
 & Yes & 5.3 years\\

DoS Attack & No & 2.9 months\\
  & Yes & 2.65 years\\
\hline
\end{tabular}
\end{table}

As observed, the impact of the preamble authentication technique on the energy consumption under normal operation conditions (no attack) is small, reducing the battery less than 4\%. 

These results confirm the fact that using the authentication preamble can reduce with 91\% the effect of an exhaustion attack, this means increasing the battery lifetime from 0.24 years to 2.65 years. 

The battery lifetime reduction due to the using of the authentication preamble under an exhaustion attack is 53.9\% (this is the worst case scenario, when in each 15s the node has to stay awake to check the preamble of the malicious message), compared with 95.6\% when not using the preamble.





\section{Conclusions}

In this article we studied the suitability of authenticated preambles to cope with exhaustion attacks in LoRaWAN networks. We demonstrate that a short 4B preamble, can incur a small energy consumption overhead of less than 4\% in operational networks while it prevents malicious attackers to significantly impact the operation of a network. This article presented results based on an industrial data logger platform used commercially for critical infrastructure control and monitoring.

\section{Acknowledgements}
This project has received funding from the European Union Horizon 2020 research and innovation programme under the Marie Sklodowska-Curie grant agreement No 675891 (SCAVENGE project) and the grant agreement No 700378 (CIPSEC). This work is also partially supported by the Spanish Ministry of Economy and the ERDF regional development fund under SINERGIA project (TEC2015-71303-R). 

%
%
%
\bibliographystyle{splncs04}
\bibliography{bibliography.bib}

\end{document}